# Low-bias electron transport properties of germanium telluride ultrathin films

Jie Liu and M. P. Anantram

*Department of Electrical Engineering, University of Washington, Seattle, WA 98195, USA*

**Abstract**

The nanometer-scale size-dependent electronic transport properties of crystalline (*c*-) and amorphous (*a*-) germanium telluride (GeTe) ultrathin films sandwiched by titanium nitride (TiN) electrodes are investigated using ab initio molecular dynamics (AIMD), density functional theory (DFT), and Green's function calculations. We find that *a*-GeTe ultrathin films scaled down to about 38 Å (12 atomic layers) still shows a band gap and the electrical conductance is mainly due to electron transport via intra-gap states. If the ultrathin films are further scaled, the *a*-GeTe band gap disappears due to overlap of the two metal induced gap states (MIGS) regions near the TiN electrodes, leading to sharp increase of *a*-GeTe conductance and significant decrease of *c*-GeTe/*a*-GeTe conductance ratio. The *c*-GeTe/*a*-GeTe conductance ratio drops below one order of magnitude if the ultrathin films are scaled below about 33 Å, making it difficult to reliably perform read operations in thin film based phase change memory devices. This overlap of the MIGS regions sets up the ultimate scaling limit of phase change memory technology. Our results suggest that the ultimate scaling limit can be pushed to even smaller size, by using phase change material (PCM) with larger amorphous phase band gap than *a*-GeTe.

## I. INTRODUCTION

While traditional silicon based integrated circuit devices (e.g. Flash and MOSFET) are becoming more and more challenging to scale, phase change material (PCM) based technologies have emerged as promising alternatives for future ultra dense memory and logic devices[1-20].

Current nanoscale silicon based memory devices suffer from problems because information (logic 0 and 1 states) is stored by the location of electrons – these electrons easily leak out of their location leading to loss of stored information. For example, electrons located in the oxide of flash memory represent a logic state; their tunneling through the oxide is becoming a problem as their dimensions scale to smaller length scales. In contrast, PCM devices do not suffer from such scaling problems because they store information by the location of atoms. Specifically, the logic 0 and 1 states are represented by the orders of magnitude difference in electrical conductance of two reversibly switchable lattice structural phases (crystalline and amorphous) of PCM thin films. This unique working mechanism offers PCM devices the potential to scale to very small thin film thickness. It is well known that decreasing the dimensions is beneficial to improve device performance. Other than increasing data density and device functionality, PCM scaling can also significantly increase device endurance[16]; decrease operation energy, current, and voltage[21-25]; and increase device speed[16,22-23]. The existing scaling studies mainly use macroscopic equations[22-23], which are based on the assumption of diffusive transport and requires experimentally calibrated phenomenological parameters as inputs. In the nanometer scale, which is our interest here, the electron transport is largely ballistic[24], and an atomistic representation is required to accurately model it. The ultimate limit to which PCM ultrathin films can be scaled but yet retain adequate crystalline (*c*-) to amorphous (*a*-) conductance ratio for reliable read operation is an open question.

In this paper, we investigate the underlying physics that determines the electrical conductance ratio between deeply scaled crystalline and amorphous thin films. We study the germanium telluride (GeTe) ultrathin films sandwiched by titanium nitride (TiN) (Fig. 1), which is the most widely-used electrode material in PCM devices due to its excellent thermal and mechanical properties. Here we choose prototypical binary PCM GeTe, instead of the most popular ternary $Ge_2Sb_2Te_5$, because binary PCM can better keep phase change properties in the ultrascaled nanostructures[3-4] probably due to its simpler stoichiometry and smaller crystalline unit cell. In this study, we use ab initio simulations, which are very computationally intensive but capable of predicting the lattice and electronic structures from first principle. As a result, we are able to provide insights into the electron transport properties at the atomistic scale, some of which are difficult to obtain using experiments. The simulation methodology is described in section II. In section III, we present the simulation results and interpret them by analyzing the transmission, local density of states, and metal induced gap states. We conclude in section IV.

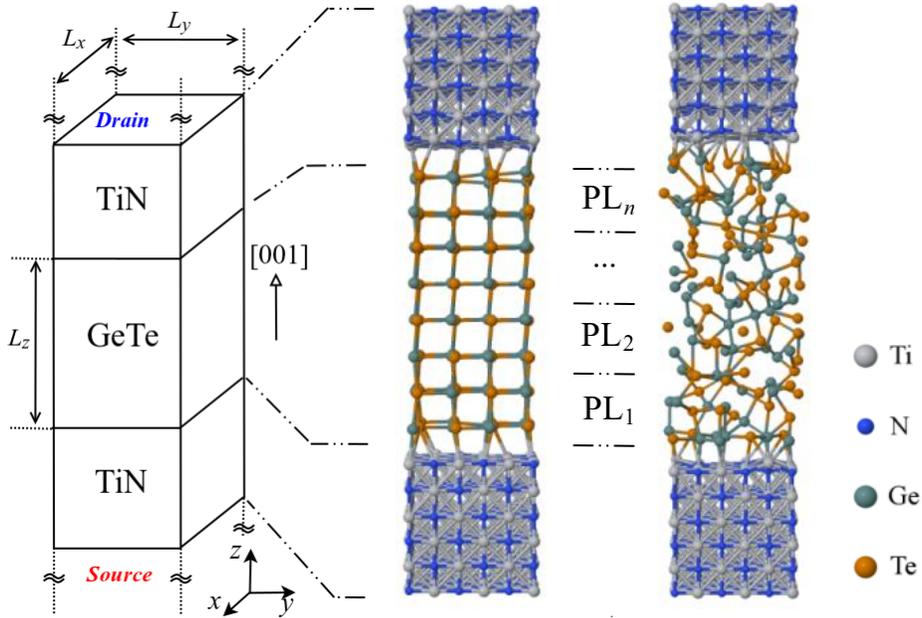

FIG. 1. (Color online) Ab initio simulation models, in which $c$-GeTe and $a$-GeTe ultrathin films with different thicknesses ($L_z$) are sandwiched by TiN electrodes. At supercell boundaries in $x$ and $y$

directions, the periodic boundary conditions (PBC) are applied. Thus, the models represent ultrathin films by infinitely repeating the supercells in *x* and *y* directions. Each principal layer (PL) contains 32 GeTe atoms in the supercell (i.e. 2 crystalline atomic layers).

## II. NUMERICAL METHODS

The ab initio simulation procedure used to create the simulation models and to obtain the electron transport properties are discussed in this section.

### A. *c*-GeTe and *a*-GeTe models

The atomic coordinates of *c*-GeTe models are obtained by relaxing the measured rhombohedral *c*-GeTe lattice structure[26] using the conjugate gradient (CG) algorithm.

To obtain the atomic coordinates of *a*-GeTe models, the *c*-GeTe is melted at 1100 K for 10 ps and then quenched to 300 K in 15 ps using ab initio molecular dynamics (AIMD) with a 5 fs timestep; then all GeTe atoms are relaxed using CG to obtain the final *a*-GeTe atomic coordinates. The AIMD and CG simulations are performed in the constant volume rectangular cuboid supercells ($L_x \times L_y \times L_z$) with the periodic boundary conditions (PBC) applied at the boundaries[10-15]. The size of *c*-GeTe (*a*-GeTe) supercell is set as $L_x=L_y=12.06$ Å and $L_z=12.06 \times N_z/4$ Å ($L_x=L_y=12.06$ Å and $L_z=13.05 \times N_z/4$ Å), such that mass density is 6.06 (5.60) g/cm$^3$ as experimentally measured[27]. Here, $N_z$ is the number of atomic layers of the ultrathin film. To generate amorphous GeTe for $N_z=8$, 10, and 12, we used brute force melt-quench AIMD simulations. For larger $N_z$ values, however, the brute-force simulation is very computationally intensive. Since the PBC is applied at the boundaries of the AIMD and CG simulation supercells, we repeat the supercell of 12 atomic layer *a*-GeTe model in the *z*-direction and take $L_z=13.05 \times N_z/4$ long region to create the *a*-GeTe models with $N_z$ atomic layers ($N_z=14$, 16, and 18).

The ab initio simulations described above give a good match to the experimentally measured pair distribution function (PDF)[28] of *a*-GeTe, as shown in Fig. 2. In the PDF results, the first (second) nearest neighbor distance is calculated to be 2.8 Å (4.1 Å), which agrees well with the measured value 2.7 Å (4.2 Å)[29]. The *a*-GeTe models show reduced density of states near Fermi level, instead of an absolute bandgap, possibly due to the intra-gap states. The *c*-GeTe model obtained using ab initio simulations shows a 0.24 eV band gap, which is also in good agreement with the experimentally measured value of 0.20 eV[30].

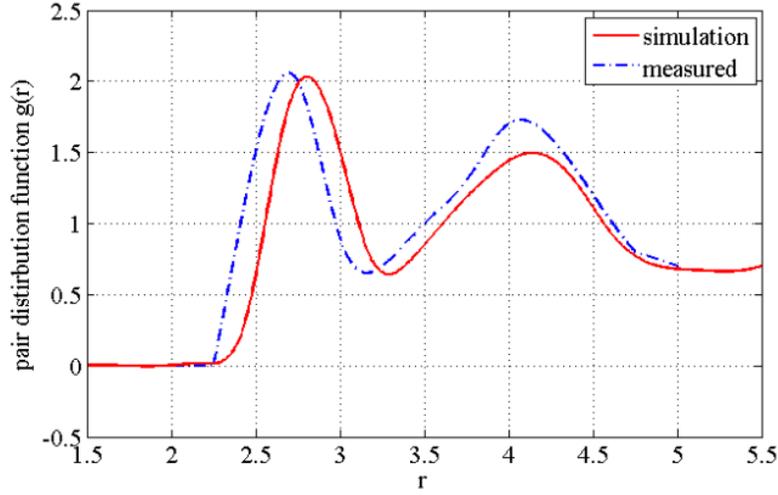

FIG. 2. (Color online) Pair distribution function (PDF) of *a*-GeTe obtained from our ab initio simulations (solid) is in good agreement with experimental data (dashed)[28].

### B. TiN-GeTe-TiN models

Then, the *c*-GeTe and *a*-GeTe models are sandwiched by TiN electrodes (Fig. 1). All Ti, N, Ge, and Te atoms in the 10 Å thick regions near each TiN-GeTe interface and the distances between TiN and GeTe are relaxed using the CG method. Finally, density functional theory (DFT) is used to obtain the Hamiltonian (**H**) and overlap (**S**) matrices of TiN-GeTe-TiN sandwich structures in the pseudo atomic orbital (PAO) representation. **H** and **S**, which contain the ab initio description of the system's electronic structure, are to be used for subsequent electron transport calculations. The CG, AIMD, and DFT algorithms used above are implemented in the SIESTA package[31].

In the AIMD, CG, and DFT simulations in this paper, Γ-point is used to sample the Brillouin Zone; the plane wave cutoff is chosen to be 100 Ry; the generalized gradient approximation (GGA) of Perdew, Burke and Ernzerhof (PBE)[32] is used to approximate the exchange-correlation energy; the double-zeta (single-zeta) plus polarization PAO is used for Ge and Te (Ti and N) atoms; and the CG relaxation convergence criteria is set as 40 meV/Å.

### C. Electron transport simulation

After the models as shown in Fig. 1 are created, the electron transport properties along the *z* direction are investigated. Since thickness of the ultrathin films of our interests here is much smaller than the electron mean free path in GeTe[24], the electrical conductance of the *c*-GeTe (*a*-GeTe) ultrathin films $G_c$ ($G_a$) is calculated by using

$$G_{c/a} = \frac{2q^2}{\hbar} \int g(\varepsilon)\, d\varepsilon \quad g(\varepsilon) = -T(\varepsilon)\frac{\partial f(\varepsilon)}{\partial \varepsilon} \quad (1)$$

in the pre-threshold voltage range, in which low bias is applied to read out the stored data. The transmission coefficient $T(\varepsilon)=\text{trace}[\Gamma_S(\varepsilon)G(\varepsilon)\Gamma_D(\varepsilon)G^\dagger(\varepsilon)]$ is computed from the retarded Green's function $G(\varepsilon)=[\varepsilon \mathbf{S}-\mathbf{H}-\Sigma_S(\varepsilon)-\Sigma_D(\varepsilon)]^{-1}$ using the recursive Green's function algorithm[33]. Here, ε is the electron energy; $f(\varepsilon)$ is the Fermi function at 300 K; $q$ is the elementary charge; $\hbar$ is the reduced Planck constant; **H** and **S** are the Hamiltonian and overlap matrices obtained using DFT; $\Sigma_{S/D}(\varepsilon)$ is the self-energy of the source/drain, which are computed using the iterative surface Green's function algorithm[34]; and $\Gamma_{S/D}(\varepsilon)=i\times[\Sigma_{S/D}(\varepsilon)-\Sigma^\dagger_{S/D}(\varepsilon)]$ is the broadening matrix of source/drain.

Here, the transmission coefficient $T(\varepsilon)$ is computed by using the Γ-point Brillouin Zone sampling. To test its precision, we compare the Γ-point sampling $T(\varepsilon)$ results against $T(\varepsilon)$ obtained by using refined $n_x \times n_y \times n_z$ Monkhorst-Pack $k$-point sampling algorithm[35]

$$T(\varepsilon) = \frac{1}{V_{BZ}} \int_{1BZ} T(\bm{k}; \varepsilon) d\bm{k} \approx \sum_i w_i T(\bm{k}_i; \varepsilon) \tag{2}$$

where $V_{BZ}$ is the area of the first Brillouin zone (1BZ) and $w_i$ is the numerical quadrature weighting coefficient corresponding to the Monkhorst-Pack sampling point $\bm{k}_i$. Among all models used in the simulation, the supercell length in the $z$-direction ranges from 49.21 Å (crystalline system when $N_z$=8) to 84.75 Å (amorphous system when $N_z$=18). As all supercell lengths in the $z$-direction are very large, $n_z$ is constantly set as 1. Supercell length in $x$ and $y$ directions are comparatively smaller ($L_x=L_y$=12.06 Å), so we choose $n_x=n_y$=1,2,3,4, in order to test the Brillouin zone sampling precision. (when $n_x=n_y$=1, $n_x \times n_y \times n_z$ Monkhorst-Pack $k$-points sampling is reduced to Γ-point sampling). Our results show that the transmission coefficient changes by less than 1% when $n_x$ and $n_y$ are increased from 1 to 4.

The local density of states (LDOS) of atom $i$ is computed by using

$$\text{LDOS}(\varepsilon, i) = -\frac{1}{\pi} \text{Im}\{\text{trace}[(\bm{G} \cdot \bm{S})_{i,i}]\} \tag{3}$$

where $\bm{G}$ is the retarded Green's function and $\bm{S}$ is the overlap matrix in the SIESTA PAO representation.

## III. RESULTS

### A. Conductance

The $L_z$-dependent $G_c$ and $G_a$ computed using Equation (1), together with the $G_c/G_a$, are shown in Fig. 3. We note that the thickest film (roughly 6 nm thick) in our ab initio simulations shows two orders of magnitude $G_c/G_a$. Interestingly, this agrees with the state-of-art scaling experiments[20], in which the thinnest Ge$_2$Sb$_2$Te$_5$ film (6 nm thick) sandwiched by TiN electrodes also shows two orders of magnitude $G_c/G_a$, though the PCM stoichiometry differs.

While the experimental $G_c/G_a$ data for sub-6 nm PCM ultrathin films is not available, we compute it as shown in Fig. 3. It can be seen that the two orders of magnitude $G_c/G_a$ can be retained even if the ultrathin film thickness is reduced down to about 40 Å (13 atomic layers). However, if the ultrathin film thickness is further reduced, the $G_c/G_a$ drops rapidly. When the thickness is reduced below about 33 Å (11 atomic layers), the $G_c/G_a$ becomes smaller than 10, which is an ON/OFF ratio value typically needed for reliable read operation in useful devices.

It worth noting that sub-20 Å PCM nanostructures still keep phase change properties[1-9], making write operations possible in ultrascaled PCM devices. Therefore, here we point out, for the first time according to our knowledge, that the aggressive scaling might be limited by the reduction of $G_c/G_a$, which makes it difficult to reliably perform read operation.

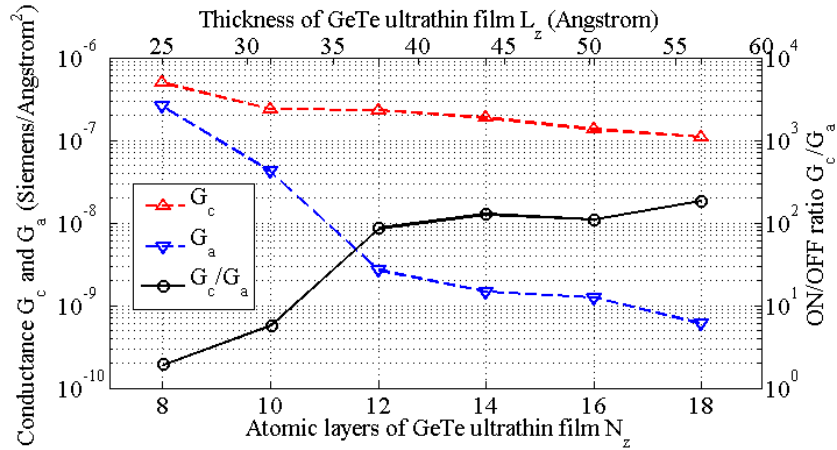

FIG. 3. (Color online) Conductance of $c$-GeTe ($G_c$) and $a$-GeTe ($G_a$) thin films, and their ratio $G_c/G_a$.

### B. Transmission

In order to understand the underlying device physics that causes the sharp reduction of $G_c/G_a$ shown in Fig. 3, we plot the $T(\varepsilon)$ and $g(\varepsilon)$ of Equation (1) in Fig. 4. The $T(\varepsilon)$ and $g(\varepsilon)$ provide energy resolved

insight – the former (latter) unravels the electron transmission probabilities (contribution to the total conductance) as a function of the electron energy ε. Here, we only present the $T(\varepsilon)$ and $g(\varepsilon)$ results for three thickness values ($N_z$=10,12,14), since the sharp reduction of $G_c/G_a$ mainly occurs in this critical thickness range.

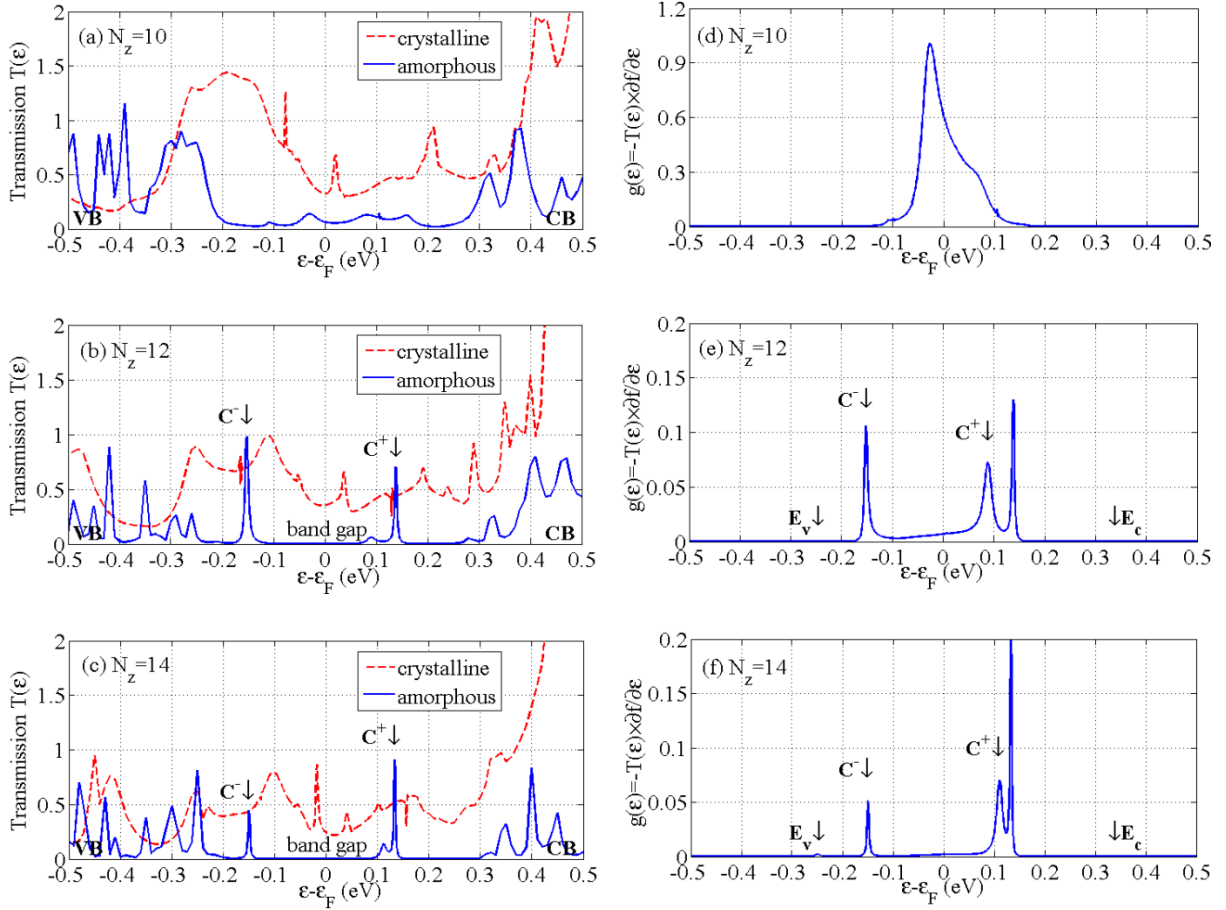

FIG. 4. (Color online) The transmission coefficient $T(\varepsilon)$ and contribution to conductance as a function of electron energy $g(\varepsilon)=-T(\varepsilon)\times \partial f(\varepsilon)/\partial \varepsilon$ of the ultrathin films. The donor-like states (C+) and acceptor-like states (C-) are clearly seen in plots of $g(\varepsilon)$ and local density of states in Fig. 5. $N_z$: number of atomic layers in the ultrathin films; CB: conduction band; VB: valence band; $E_c$: conduction band edge; $E_v$: valence band edge; and $\varepsilon_F$: Fermi level.

Physically, the *c*-GeTe and *a*-GeTe ultrathin films, which are sandwiched by metal TiN electrode as shown in Fig. 1, form potential barriers for electron transport during read operation. When the lattice structure is changed between the amorphous phase and the crystalline phase, the electronic structure of the barrier is altered, causing the change of the electron transmission behavior as shown in Fig. 4(a)-(c).

It is obvious that near the Fermi level $\varepsilon_F$, the amorphous phase $T(\varepsilon)$ is always smaller than the crystalline phase $T(\varepsilon)$, making the conductance of the amorphous phase $G_a$ smaller than that of the crystalline phase $G_c$ (Fig. 3). If the ultrathin films are scaled down to 12 atomic layers, the amorphous phase still exhibits clear band gaps (Fig. 4(b)-(c)) and the electrical conductance is mainly due to the electron transport via intra-gap donor-like and acceptor-like states (Fig. 4(e)-(f)). However, if the ultrathin film is further scaled, the amorphous phase band gap disappears (Fig. 4(a)), leading to both quantitatively significant increase and qualitatively sharp change of $g(\varepsilon)$ (Fig. 4(d)). Now the conductance is no longer determined by the distinct donor-like states and acceptor-like states, but dominated by transport near the Fermi level of the metal electrodes.

In contrast to the significant change of electron transport behavior when the amorphous phase film is aggressively scaled, we observe only a slight quantitative increase of electron transmission probabilities when the crystalline phase ultrathin film is scaled to the same dimensions (Fig. 4(a)-(c)). The crystalline films in the size scale of our interests here show no band gap, and the conductance is always dominated by electron transport near the Fermi level. As a consequence, when the ultrathin films are aggressively scaled, $G_a$ increases significantly but $G_c$ increases only slightly (Fig. 3). Therefore, the decrease of $G_c/G_a$ is mainly caused by the rapid increase of $G_a$ with decreasing film thickness.

### C. Local density of states

The difference in scaling behavior between *c*-GeTe and *a*-GeTe ultrathin films evokes our interest to inspect their electronic structure difference, in order to further understand the governing device physics.

In Fig. 5, we present the LDOS at the most important thickness point ($N_z=12$), since immediately beyond this length scale, the $G_c/G_a$ decreases rapidly during scaling (Fig. 3).

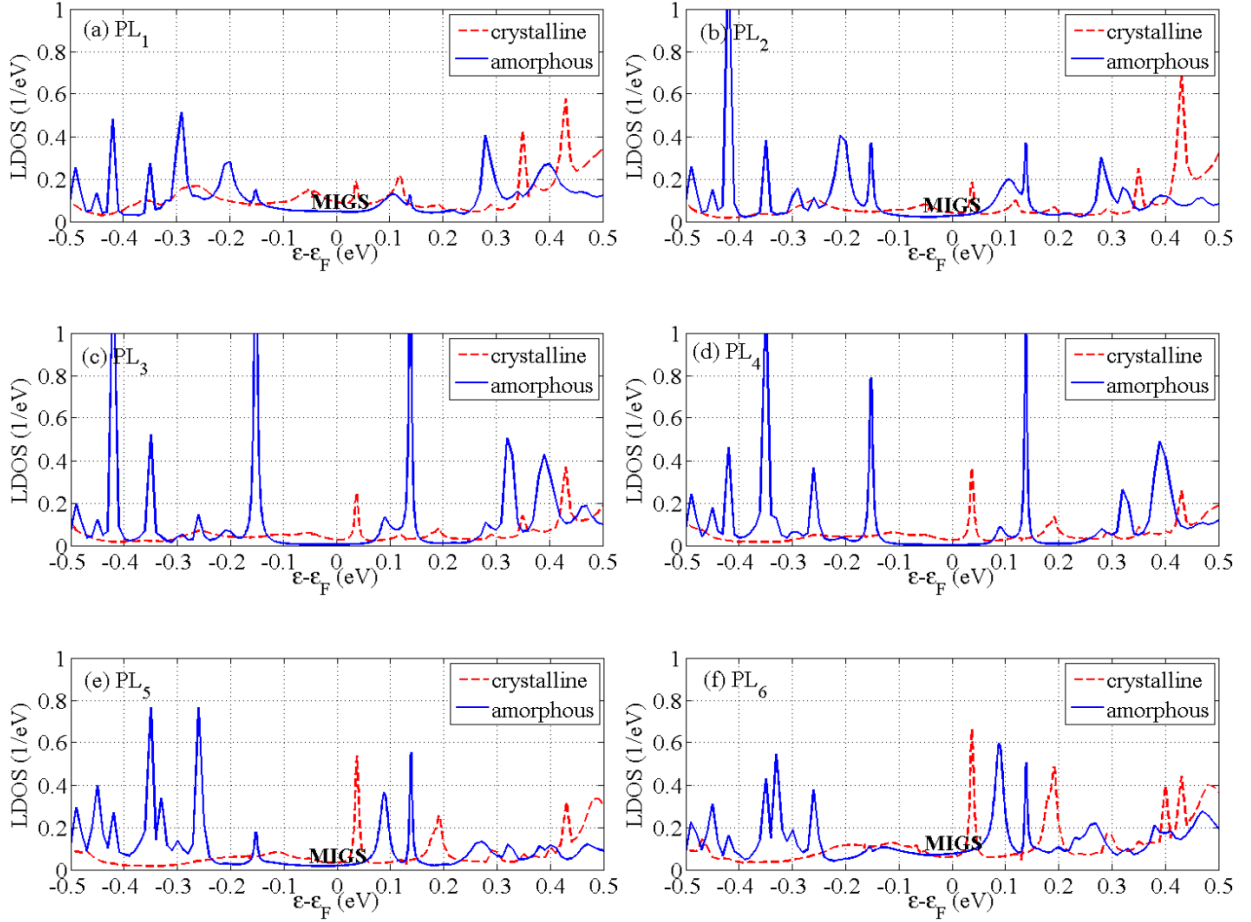

FIG. 5. (Color online) Local Density of State (LDOS) of different principal layers ($PL_n$, as defined in Figure 1) in the GeTe ultrathin film ($N_z=12$). MIGS: metal induced gap states in amorphous phase

In the discussion here, a principal layer (PL) is defined as 32 Ge and Te atoms, as shown in Fig. 1 (for crystalline phase, one PL means two atomic layers of $c$-GeTe). The ultrathin film with $N_z=12$ consists of six PLs stacking in the $z$-direction from source to drain. Each of the six PLs is labeled $PL_n$ where $n$ varies from 1 to 6. It can be seen from Fig. 5 that in the middle region of $a$-GeTe ultrathin films, which is far away from the metal electrodes ($PL_3$ and $PL_4$), the LDOS is very small near the Fermi level, and there is a clear band gap with pronounced intra-gap donor-like and acceptor-like states (Fig. 5(c)-(d)). These states are responsible for the conduction in Fig 4(b). In the $a$-GeTe region that is spatially close to TiN electrodes ($PL_1$, $PL_2$, $PL_5$, and $PL_6$), the band gap disappears and the LDOS is large near the Fermi level (Fig. 5(a)-(b) and Fig. 5(e)-(f)) due to the hybridization with the electronic states inside the metal electrodes. These metal induced gap states (MIGS) evanescently penetrate two PLs into the $a$-GeTe (Fig. 5(a)-(b) and Fig. 5(e)-(f)) and decay in the deeper $a$-GeTe region (Fig. 5(c)-(d)).

In comparison, the MIGS penetrate all the way across the ultrathin film in c-GeTe thin films (Fig. 5). As a result, near the Fermi level, the $c$-GeTe LDOS is much larger than the $a$-GeTe LDOS in the region far away from the TiN electrodes (Fig. 5(c)-(d)). This LDOS magnitude difference of the two phases explains why the $T(\varepsilon)$ of $c$-GeTe is larger than that of $a$-GeTe, and why $G_c$ is always larger than $G_a$ (Fig. 3).

### D. Role of metal induced gap states

It worth mentioning that, in the two regions ($PL_1$ and $PL_2$; $PL_5$ and $PL_6$) close to the TiN electrodes, the $a$-GeTe LDOS is in general comparable to, though slightly smaller than, the $c$-GeTe LDOS near the Fermi level (Fig. 5(a)-(b) and Fig. 5(e)-(f)). If the ultrathin films are thick enough ($N_z>12$), these two MIGS-controlled high-LDOS conductive regions are spatially separated by a MIGS-free low-LDOS resistive region ($PL_3$ and $PL_4$) when the PCM state is amorphous, making the amorphous phase much more resistive than the crystalline phase. But if the ultrathin films are aggressively scaled ($N_z<10$), these two regions become closer, bridging a path for electrons to transport through the amorphous phase. As a consequence, in the amorphous phase transmission probability (Fig. 4(a)), tunneling via MIGS near the Fermi level (Fig. 4(d)), and conductance $G_a$ (Fig. 3) increases significantly, leading to the rapid decrease of $G_c/G_a$.

As mentioned above, we observed that the MIGS penetration depth in *c*-GeTe is larger than that in *a*-GeTe. This phenomenon can be attributed to the well known fact that *c*-GeTe band gap is smaller than *a*-GeTe band gap and in general MIGS penetrate deeper into the semiconductors with smaller band gap at the metal-semiconductor interface. The above analysis on the role of MIGS during scaling indicates that if the PCM with larger amorphous phase band gap is used, the MIGS penetration depth can be reduced and the two MIGS-controlled regions can be better separated. Therefore, it might be promising to push the ultimate scaling limit of PCM to even smaller dimension by using PCM which has a larger amorphous phase band gap, or by applying other techniques that can effectively reduce the MIGS penetration depth. We also note that the larger amorphous band gap will lead to increased threshold voltage required for the amorphous to crystalline transformation[24]. This is a typical example where there is a need to strike a balance between two competing device performance metrics. Although it has been shown that ultrathin films can be scaled to at least 33 Å from the viewpoint of keeping adequate $G_c/G_a$ for reliable read operation, we note that the scaling scenario can also be shaped by many other technical issues and device performance metrics, e.g. crystallization temperature, operation energy, device endurance, and fabrication cost. Their roles in determining the PCM scaling scenario deserve future research efforts. Finally, we remark that while using DFT can result in inaccurate band gaps, we hope that the qualitative findings in our results will help understand scaling properties of the electron transport in ultrascaled PCM nanostructures and establish larger crystalline to amorphous conductance ratio, $G_c/G_a$.

## IV. CONCLUSION

In conclusion, we applied ab initio methods to investigate the electron transport properties of the GeTe ultrathin films in both its amorphous and crystalline phases, sandwiched by TiN electrodes. The amorphous phase was generated by a melt quench simulation of the crystalline material using ab initio molecular dynamics. We find that as the film thickness decreases, the bandgap of the amorphous material disappears due to the overlap of two metal induced gap state (MIGS) regions near the TiN electrodes. We also find that when the *a*-GeTe ultrathin films scale down to about 38 Å (12 atomic layers), the thin film still shows a band gap, though the electrical conductance is mainly due to electron transport via intra-gap states. Futher decrease of film thickness leads to the appearance of large MIGS near the Fermi energy, and a concomitant large decrease of *c*-GeTe/*a*-GeTe conductance ratio. The *c*-GeTe/*a*-GeTe conductance ratio drops below one order of magnitude if the ultrathin films are scaled below about 33 Å. The overlap of the MIGS regions sets up the ultimate scaling limit of phase change memory. Our results suggest that the ultimate scaling limit can be pushed to even smaller sizes by using phase change material with a larger amorphous phase band gap than *a*-GeTe. Our analysis also indicates that the down scaling of phase change material thickness might be limited by requirements of the read operation, instead of the write operations as widely believed[1-9]. We show that the metal induced gap states (MIGS) in the amorphous phase play a pivotal role in determining the smallest film thickness. We pointed out that by reducing MIGS in amorphous phase, the ultimate scaling limit of PCM can be further reduced. These insights are of importance to the scalability of future PCM devices.

## ACKNOWLEDGEMENTS

This work was supported by U.S. National Science Foundation (NSF) under Grant Award 1006182. This work used the Extreme Science and Engineering Discovery Environment (XSEDE), which is supported by National Science Foundation grant number OCI-1053575. This work was facilitated through the use of advanced computational, storage, and networking infrastructure provided by the Hyak supercomputer system, supported in part by the University of Washington's eScience Institute. We acknowledge J. Akola and R.O. Jones (for discussion about AIMD simulations) and Xu Xu (for discussion about *k*-point sampling). A part of this work was done during the first author's summer intern in Micron Technology, Inc, and we acknowledge the help and support of G. Sandhu, R. Meade, and the Micron Technology, Inc.

---

[1]. R. E. Simpson, M. Krbal, P. Fons, A. V. Kolobov, J. Tominaga, T. Uruga, and H. Tanida,. Nano Letters. **10**, 414 (2010).

[2]. M. A. Caldwell, S. Raoux, R. Y. Wang, H. P. Wong, and D. J. Milliron, J. Mater. Chem. **20**, 1285 (2010).

[3]. S. Raoux, J. L. Jordan-Sweet, and A. J. Kellock, J. Appl. Phys. **103**, 114310 (2008).

[4]. S. Raoux, C. T. Rettner, J. L. Jordan-Sweet, V. R. Deline, J. B. Philipp, and H. L. Lung, Proc. of Euro. Symp. on Phase Change and Ovonic Science, 127, (2006).

[5]. H. P. Wong, S. Kim, B. Lee, M. A. Caldwell, J. L. Liang, Y. Wu, R. G. D. Jeyasingh, and S. M. Yu, 10th IEEE International Conference on Solid-State and Integrated Circuit Technology, (2010).

[6]. H. P. Wong, S. Raoux, S. Kim, J. L. Liang, J. P. Reifenberg, B. Rajendran, M. Asheghi, and K.E. Goodson, Proc. of IEEE, **98**, 2201-2227 (2010).

[7]. G. W. Burr, M. J. Breitwisch, M. Franceschini, D. Garetto, K. Gopalakrishnan, B. Jackson, B. Kurdi, C. Lam, L. A. Lastras, A. Padilla, B. Rajendran, S. Raoux, and R. S. Shenoy, J. Vac. Sci. Technol. B, **28**, 223 (2010).

[8]. S. Raoux, G. W. Burr, M. J. Breitwisch, C. T. Rettner, Y. C. Chen, R. M. Shelby, M. Salinga, D. Krebs, S. H. Chen, H. L. Lung, and C. H. Lam, IBM J. Res. & Dev. **52**, 465-479 (2008).

[9]. S. Raoux, M. Wuttig, Eds., Phase Change Materials Science and Applications, Springer, New York, (2009).

[10]. J. Akola, J. Larrucea, and R. O. Jones, Phys. Rev. B **83**, 094113 (2011).

[11]. J. Akola, R. O. Jones, S. Kohara, S. Kimura, K. Kobayashi, M. Takata, T. Matsunaga, R. Kojima, and N. Yamada, Phys. Rev. B **80**, 020201 (2009).

[12]. J. Akola, and R. O. Jones, Phys. Rev. B **79**, 134118 (2009).

[13]. J. Akola and R. O. Jones, Phys. Rev. B **76**, 235201 (2007).

[14]. T. H. Lee and S. R. Elliott, Phys. Rev. B **84**, 094124 (2011).

[15]. T. H. Lee and S. R. Elliott, Phys. Rev. Lett. **107**, 145702 (2011).

[16]. W. J. Wang, D. Loke, L. P. Shi, R. Zhao, H. X. Yang, L. T. Law, L. T. Ng, K. G. Lim, Y. C. Yeo, T. C. Chong, and A. L. Lacaita, Nature Sci. Rep. **2**, 1, (2011).

[17]. F. Xiong, A. Liao, D. Estrada, andE. Pop, Science, 332, 568 (2011).

[18]. S. W. Nam, H. S. Chung, Y. C. Lo, L. Qi, J. Li, Y. Lu, A. T. C. Johnson, Y. Jung, P. Nukala, and R. Agarwal, Science **336**, 1561 (2012).

[19]. D. Loke, T. H. Lee, W. J. Wang, L. P. Shi, R. Zhao, Y. C. Yeo, T. C. Chong, and S. R. Elliott, Science, **336**, 1566 (2012).

[20]. S. B. Kim,; B. J. Bae, Y. Zhang, R. G. D. Jeyasingh, Y. Kim, I. G. Baek, S. Park, S. W. Nam, and H. P. Wong, IEEE Trans. on Elect. Dev. **58**, 1483 (2011).

[21]. A. Pirovano, A. L. Lacaita, A. Benvenuti, F. Pellizzer, S. Hudgens, and R. Bez, Tech. Digest of IEEE Intl. Elect. Dev. Meet. 29.6.1 (2003) .

[22]. S. Kim and H. P. Wong, IEEE Electr. Dev. Lett. **28**, 697 (2007).

[23]. J. Liu, B. Yu, and M. P. Anantram, IEEE Elect. Dev. Lett. **32**, 1340 (2011).

[24]. Yu, D.; Brittman, S.; Lee, J. S.; Falk, A. L.; Park, H., Nano Letters **8**, 3429 (2008).

[25]. D. Krebs, S. Raoux, C. T. Rettner, G. W. Burr, M. Salinga, and M. Wuttig, Appl. Phys. Lett. **95**, 082101 (2009).

[26]. T. Chattopadhyay, J. X. Boucherle, and H. G. Schnering, J. of Phys. C. **20**, 1431 (1987).

[27]. T. Nonaka,; G. Ohbayashi, Y. Toriumi, Y. Mori, and H. Hashimoto, Thin Solid Films **370**, 258 (2000).

[28]. G. E. Ghezzi, Appl. Phys. Lett. **99**, 151906 (2011).

[29]. D. B. Dove, M. B. Heritage, K. L. Chopra, S. K. Bahl, Appl. Phys. Lett. **16**, 138 (1970).

[30]. L. L. Chang, P. J. Stiles, and L. Esaki, IBM J. of Res. and Dev. **10**, 484 (1966).

[31]. J. M. Soler, E. Artacho, J. D. Gale, A. Garca, J. Junquera, P. Ordejon, and D.Sanchez-Portal, J. Phys.: Condens. Matter, **14**, 2745 (2002).

[32]. J. P. Perdew, K. Burke, and M. Ernzerhof, Phys. Rev. Lett. **77**, 3865 (1996).

[33]. R. Lake, G. Klimeck, R. C. Bowen, D. Jovanovic, J. Appl. Phys. **81**, 7845 (1997).

[34]. M. P. L. Sancho, J. M. L. Sancho, and J. Rubio, J. Appl. Phys. F. **15**, 851 (1985).

[35]. H. J. Monkhorst and J. D. Pack, Phys. Rev. B, **13**, 5188 (1976).